\documentclass{ws-procs9x6}

\def\al{\alpha}
\def\be{\beta}
\def\ga{\gamma}
\def\de{\delta}

\def\ta{\tau}

\def\fr#1#2{{{#1} \over {#2}}}

\def\frac#1#2{{\textstyle{{#1}\over {#2}}}}

\def\lsim{\mathrel{\rlap{\lower4pt\hbox{\hskip1pt$\sim$}}
    \raise1pt\hbox{$<$}}}
\def\gsim{\mathrel{\rlap{\lower4pt\hbox{\hskip1pt$\sim$}}
    \raise1pt\hbox{$>$}}}
\def\sqr#1#2{{\vcenter{\vbox{\hrule height.#2pt
         \hbox{\vrule width.#2pt height#1pt \kern#1pt
         \vrule width.#2pt}
         \hrule height.#2pt}}}}

\def\lrpartial{\raise 1pt\hbox{$\stackrel\leftrightarrow\partial$}}

\def\etal{{\it et al.}}

\newcommand{\beq}{\begin{equation}}
\newcommand{\eeq}{\end{equation}}
\newcommand{\bea}{\begin{eqnarray}}
\newcommand{\eea}{\end{eqnarray}}
\newcommand{\rf}[1]{(\ref{#1})}

\begin{document}

\title{Lorentz violation 
as a quantum-gravity signature\footnote{\uppercase{F}unded in part 
by the \uppercase{C}entro \uppercase{M}ultidisciplinar 
de \uppercase{A}strof\'{\i}sica (\uppercase{CENTRA})
and by the \uppercase{F}unda\c{c}\~ao para a 
\uppercase{C}i\^encia e a \uppercase{T}ecnologia 
(\uppercase{P}ortugal)
under grant \uppercase{POCTI}/\uppercase{FNU}/49529/2002.}}

\author{Ralf Lehnert}

\address{CENTRA, Departamento de F\'{\i}sica,
Universidade do Algarve\\
8000-117 Faro, Portugal\\
E-mail: rlehnert@ualg.pt}

\maketitle

\abstracts{Many theoretical approaches to quantum gravity 
predict the breakdown of Lorentz symmetry at Planck energies. 
Kinematical cosmic-ray studies are 
a sensitive tool in the search for such effects. 
This talk discusses the construction 
of test dispersion relations for such analyses.
}

\section{Introduction}

One of the most fundamental problems in present-day physics
concerns a quantum theory of gravitation.
Such a theory is believed to be associated
with the Planck scale $M_{Pl}$,
so that quantum-gravity signatures
are likely to be suppressed
by one or more powers of $M_{Pl}$
at currently attainable energies.
The expected minuscule size of potential effects
and the absence of a fully realistic and satisfactory underlying theory
make quantum-gravity phenomenology particularly challenging.
A practical approach to overcome this obstacle
is to consider the breaking of symmetries
that hold exactly in our current fundamental laws,
might be violated in approaches to quantum gravity,
and can be tested with ultrahigh precision.

Lorentz- and CPT-symmetry breakdown
offers a promising tool
in this line of investigation.\cite{cpt01}
An effective-field-theory framework
for the description of such effects
at present experimental energy scales
is provided by the Standard-Model Extension (SME).\cite{ck,grav,kl01}
The SME coefficients
parametrizing Lorentz and CPT violation
can arise in numerous approaches to quantum gravity
including strings,\cite{kps}
spacetime-foam models,\cite{lqg,kp03,klink}
noncommutative field theory,\cite{ncft}
cosmologically varying scalars,\cite{vc,aclm}
random-dynamics models,\cite{rd}
multiverses,\cite{mv}
and brane-world scenarios.\cite{bws}

The flat-spacetime limit of the SME
has provided the basis for
numerous experimental and theoretical investigations
of Lorentz and CPT breakdown,
many of which place tight constraints on SME parameters.
Examples of such analyses
include ones involving
mesons,\cite{hadronexpt,kpo,hadronth,ak}
baryons,\cite{ccexpt,spaceexpt,cane}
electrons,\cite{eexpt,eexpt2,eexpt3}
photons,\cite{vc,photonexpt,photonth,cavexpt,km}
and muons.\cite{muons}
We also remark
that neutrino-oscillation experiments
offer the potential for discovery.\cite{ck,neutrinos,nulong}

The quadratic sector of the SME
determines the one-particle dispersion relations,
which typically exhibit 
Lorentz-violating modifications,\cite{photonexpt,ck,kl01}
as expected.
It follows
that particle-reaction kinematics
is altered relative to the conventional case
leading to potentially observable shifts
in certain reaction thresholds.
At energies approaching the Planck scale,
such Lorentz-violating effects
might be more pronounced,
so that kinematical investigations
involving ultrahigh-energy cosmic rays (UHECR)
can provide a high-sensitivity laboratory
for Lorentz-violation searches.

Many recent quantum-gravity investigations
outside the context of the SME
have exploited the general idea
to constrain deviations from conventional dispersion relations.\cite{dr,rl03}
Because of the absence of a complete underlying theory,
the form of the considered Lorentz-breaking dispersion-relation modifications
appears somewhat arbitrary in some of these studies.
This talk addresses the question
as to whether at least fundamental physics principles
that are expected to remain valid in an underlying theory
can yield meaningful restrictions
on possible dispersion-relation modifications.
We primarily consider the principle of coordinate independence
and the requirement of compatibility
with an effective dynamical framework.
We explain
why these two conditions
should continue to hold in the presence of Lorentz violation
regardless of the details of the Planck-scale theory
and discuss the ensuing constraints on dispersion relations.
Throughout we assume translational invariance
and the associated energy--momentum conservation.

The outline of this talk is as follows.
Section \ref{ci}
discusses the need for coordinate independence
and its consequences.
In Sec.\ \ref{eft},
we comment on the compatibility of the modified dispersion relations
with dynamics.
Some subtle issues regarding the applicability
of certain conventional approximations
in the Lorentz-violating context
are addressed in Sec.\ \ref{appr}.

\section{Coordinate independence}
\label{ci}

Although coordinate systems 
are one of the most important mathematical tools 
in physics, 
they do not possess physical reality: 
the description of a process must remain independent 
of the choice of coordinates. 
This allows, for example, 
different observers,
each equipped with a different reference frame, 
to relate their observations or predictions 
concerning a given physical system. 
The principle of coordinate independence 
is therefore also called observer invariance. 
Mathematically, 
observer invariance is usually implemented 
by choosing a spacetime-manifold description
for physical events, 
tensors or spinors 
for the representation of observables, 
and covariant equations as the laws of physics. 

It is a common misconception 
that Lorentz breaking 
implies the loss of coordinate independence. 
Such a model would inhibit meaningful physical predictions, 
so that it is necessary to adhere to observer invariance 
also in the presence of Lorentz violation. 
This is by no means a contradiction. 
Consider, for instance, 
the conventional motion 
of a classical point particle of mass $m$ and charge $q$ 
in an external electromagnetic field $F^{\mu\nu}$ 
described by 
\beq
m\:\fr{dv^{\mu}}{d\ta}=qF^{\mu\nu}v_{\nu}\; .
\label{example}
\eeq
Here, $\ta$ denotes the proper time
and $v^{\mu}$ the four-velocity of the particle. 
Note that invariance under rotations of the particle's trajectory 
is broken by the external nondynamical $F^{\mu\nu}$ 
resulting in the nonconservation of the charge's angular momentum. 
However, 
Eq.\ \rf{example} is a tensor equation 
valid in all coordinate systems 
maintaining coordinate independence. 
This example also illustrates 
that transformations of localized dynamical particles and fields 
(with nondynamical global backgrounds held fixed) 
must be clearly distinguished from transformations of the coordinates. 
The former, 
also called particle transformations, 
are no longer associated 
with a symmetry of the model. 

In what follows, 
we consider models that exhibit particle Lorentz violation 
while maintaining coordinate independence. 
In many respects, 
the physical situation is conceptually similar 
to the external-field case 
discussed in the previous paragraph. 
However, 
in the above example  
the $F^{\mu\nu}$ background is a local electromagnetic field 
generated by other charge and current distributions 
that can in principle be controlled. 
In the present Lorentz-violating context, 
such a background is a global property of the effective vacuum 
outside of experimental control. 

We continue our study in a local Minkowski frame. 
Lorentz-violating dispersion relations 
are usually taken to be of the form 
\beq
{E_0}^2-{\vec{p}}^{\:2}=m^2+\de f(E,\vec{p})\; ,
\label{lvdr}
\eeq
where $p^{\mu}=(E,\vec{p})$ and $m$
are the particle's respective four-momentum and mass, 
and $\de f(E,\vec{p})$ 
describes the Lorentz violation. 
With our above consideration, 
the correction $\de f$ 
needs to be observer Lorentz invariant, 
i.e., 
it must transform as a scalar 
under coordinate changes. 
To obtain a general form of $\de f$, 
we impose some further mild conditions. 
First, 
for small momenta\footnote{Here, $\vec{p}$ refers to the components 
in any frame in which the Earth moves nonrelativistically.} 
$|\vec{p}|\ll M_{Pl}$ 
we want to recover the usual dispersion relation with $\de f=0$. 
Second, 
we want to avoid potential nonlocalities 
that could arise through the presence of nonpolynomial functions. 
This yields the ansatz 
\beq
\de f(E,\vec{p})=
\sum_{n\ge 1}\hspace{4mm}
\overbrace{\hspace{-3.5mm}T_{(n)}^{\;\al\be\;\cdots\;\;}}^{n\; \rm indices}
\hspace{-1mm}\underbrace{\hspace{.5mm}p_{\al}p_{\be}\;\cdots\;}
_{n\; \rm factors} \;\; .
\label{ansatz}
\eeq
Here, 
$T_{(n)}^{\;\al\be\;\cdots\;}$ is a constant tensor of rank $n$ 
describing particle Lorentz violation. 
All the tensor indices $\al,\be\,\ldots$ 
are distinct, 
and each one is properly contracted with a momentum factor, 
so that each term in the sum is coordinate independent. 

This ansatz has various consequences,\cite{rl03} 
two of which we mention next: 
when viewed as a polynomial in $E$, 
the modified dispersion relation \rf{lvdr} 
will in general lift the usual degeneracy 
between particle, antiparticle, 
and possible spin-type states. 
Another implication concerns cases 
with imposed rotational invariance. 
Then, 
the general ansatz \rf{ansatz} 
does not contain odd powers of $|\vec{p}|$.\cite{rl03}

\section{Effective Field Theory}
\label{eft}

Kinematical considerations can impose 
powerful restrictions on particle reactions. 
However, dispersion-relation constraints 
can be masked by other effects. 
For example, 
a high-energy reaction expected to be suppressed 
by a modified dispersion relation, 
could also be prevented 
by a novel symmetry. 
Similarly, 
the presence at high energies 
of a reaction kinematically forbidden at low energies 
might perhaps be explained 
by additional channels 
due to  new undetected particles
or the loss of 
low-energy symmetries. 
In addition, 
models of both acceleration mechanisms for UHECRs 
and atmospheric shower development 
involve conventional dynamics. 

We see that the implementation 
of general dynamical features 
both appears necessary 
for a complete description of UHECR physics 
and can increase the scope 
of cosmic-ray analyses. 
At the same time, 
it may introduce a certain degree of framework dependence. 
However, 
implementing dynamics 
is tightly constrained 
by the requirement 
that known physics must be recovered 
in certain limits. 
In what follows, 
we argue that effective field theory (EFT)
is a sensible and general approach 
for such efforts. 

EFTs 
have proved to be extremely successful 
in describing diverse physical systems 
at atomic, nuclear, and elementary-particle scales. 
Note in particular, 
that EFT is flexible enough 
for applications 
involving discrete condensed-matter backgrounds, 
situations analogous to those in some quantum-gravity approaches. 
Moreover, 
the usual Standard Model itself 
is normally viewed 
as an EFT 
approximating more fundamental physics. 

The construction of a suitable EFT 
can employ a philosophy 
paralleling that of the dispersion-relation approach 
in the previous section 
with the more powerful idea 
of proceeding at the Lagrangian level: 
one adds Lorentz-breaking terms $\de {\mathcal L}$ 
to the usual Standard-Model Lagrangian ${\mathcal L}_{\rm SM}$  
\beq
{\mathcal L}_{\rm SME}={\mathcal L}_{\rm SM}+\de {\mathcal L}\; , 
\label{sme}
\eeq
where ${\mathcal L}_{\rm SME}$ denotes the EFT Lagrangian. 
The correction $\de {\mathcal L}$ 
is formed by contracting Standard-Model field operators 
of unrestricted dimensionality 
with Lorentz-breaking tensorial parameters 
(analogous to the $T_{(n)}^{\;\al\be\;\cdots\;}$ in Eq.\ \rf{ansatz}) 
yielding observer Lorentz scalars. 
This general EFT for Lorentz violation 
is the SME mentioned in the introduction. 
Its apparent generality 
makes it difficult 
and perhaps even impossible 
to find some other effective theory for Lorentz breaking 
containing the Standard Model 
with dynamics significantly different from the SME.

\section{Additional considerations}
\label{appr}

In this section, 
we point out 
that care is required 
when approximating Lorentz-violating dispersion relations. 
This is best demonstrated 
with a specific example. 

Consider the modified dispersion relation 
\beq
{E}^2-\vec{p}^{\:2}=m^2+\fr{E\vec{p}^{\:2}}{M}\; ,
\label{foamdr}
\eeq
where $M$ is some high-energy scale, 
such as $M_{Pl}$. 
The usual quantum-field reinterpretation 
of the negative-energy solutions 
yields the respective 
exact particle and antiparticle energies 
$E_+$ and $E_-$: 
\beq 
E_{\pm}(\vec{p})=\sqrt{\fr{\vec{p}^{\:4}}{4M^2}+\vec{p}^{\:2}+m^2} 
\; \pm \fr{\vec{p}^{\:2}}{2M}\; . 
\label{eigenen2} 
\eeq 
If one instead employs the ultrarelativistic approximation 
$E\simeq|\vec{p}|$ in the correction term, 
as is sometimes done in such kinematical studies, 
one obtains $\de f\simeq|\vec{p}|^3/M$. 
After the reinterpretation, we now have 
\beq 
E_{\pm}(\vec{p})=\sqrt{\fr{|\vec{p}|^{\,3}}{M}+\vec{p}^{\:2}+m^2}\; . 
\label{eigenen} 
\eeq 

Next, 
we look at photon decay 
into an electron--positron pair\footnote{This process 
is kinematically forbidden 
in conventional physics.} $\ga\rightarrow e^++e^-$ 
assuming a dispersion relation of the type \rf{foamdr} 
for both the leptons and the photon. 
For each particle, 
we take $M=-M_{Pl}$, 
and for the photon we set $m=0$. 
Using the exact expression for the particle energies \rf{eigenen2}, 
one has to consider two distinct incoming photon states $\ga_+$ and $\ga_-$, 
where the subscripts 
correspond to those of the particle energy in Eq.\ \rf{eigenen2}. 
One can then show\cite{rl03} 
that the decay $\ga_{-}\rightarrow e^++e^-$ is allowed 
above a certain threshold. 
If the observed value $m=0.511$ MeV for the electron and positron masses 
is used, 
the numerically determined threshold value 
for the incoming photon three-momentum 
is $|\vec{p}_{\rm min}|\simeq 7.21$ TeV. 
If one instead employs the approximate particle energies \rf{eigenen}, 
photon decay is forbidden 
throughout the validity range of Eq.\ \rf{eigenen}. 

Considering the TeV threshold scale for this decay 
and the Planck suppression of the correction $\de f$, 
the ultrarelativistic approximation 
is indeed excellent 
if one is interested in the particle energy only. 
However, 
threshold analyses are based on exact energy--momentum conservation 
and can thus be sensitive 
to the slightest deviations. 
Even in a conventional photon decay, 
the ultrarelativistic approximation 
renders the lepton momenta lightlike, 
which seemingly permits the decay in forward direction. 
In the present case, 
$E\simeq|\vec{p}|$ introduces 
an additional degeneracy into the problem. 
As a result, 
the approximate solution 
is spacelike, 
whereas the exact expression \rf{eigenen2} 
determines both a timelike and a spacelike branch.
It is the presence of the timelike momenta
that permits the decay.


\end{document}